\documentclass{PoS}

\newcommand{\gsim}{\;\rlap{\lower 3.5 pt \hbox{$\mathchar \sim$}} \raise 1pt
 \hbox {$>$}\;}
\newcommand{\lsim}{\;\rlap{\lower 3.5 pt \hbox{$\mathchar \sim$}} \raise 1pt
 \hbox {$<$}\;}

\title{Higgs boson pair production at the LHC: top-quark mass effects at
next-to-leading order}

\ShortTitle{Higgs boson pair production at the LHC: top-quark mass effects
at NLO}

\author{
  Jonathan Grigo\\
  Karlsruhe Institute of Technology (KIT), Karlsruhe, Germany\\
  E-mail: \email{jonathan.grigo@kit.edu}}

\author{
  Jens Hoff\\
  Karlsruhe Institute of Technology (KIT), Karlsruhe, Germany\\
  E-mail: \email{jens.hoff@kit.edu}}

\author{Kirill Melnikov\\
        Johns Hopkins University, Baltimore, MD, USA\\
        E-mail: \email{melnikov@pha.jhu.edu}}

\author{
  \speaker{Matthias Steinhauser}%
  \\
  Karlsruhe Institute of Technology (KIT), Karlsruhe, Germany\\
  E-mail: \email{matthias.steinhauser@kit.edu}}

\abstract{Higgs boson pair production is considered at next-to-leading order
  with special emphasis on the effect of a finite top quark mass. It is shown
  that, unlike for single-Higgs boson production, power-suppressed corrections
  are numerically important.}

\FullConference{11th International Symposium on Radiative Corrections
  (Applications of Quantum Field Theory to Phenomenology) \\
  22-27 September 2013\\
  Lumley Castle Hotel, Durham, UK}

\begin{document}

\section{Introduction}

Since the discovery of a Higgs boson at the CERN Large Hadron Collider
(LHC)~\cite{atlasd,cmsd} 
there has been an enormous activity aiming for the determination of the
properties of the new particle. Among them are the spin, 
the couplings to fermions and bosons and the self-coupling. 
The latter is part of the Higgs boson potential which can be written
as
\begin{eqnarray}
  V_{\rm H} = \frac{1}{2} m_H^2\, H^2 + \lambda\, v\, H^3 +
  \frac{1}{4}\,\lambda\, H^4
  \,,
\end{eqnarray}
where $H$ is the physical Higgs boson field and $v$ is the vacuum expectation
value. In the Standard
Model (SM) we have $\lambda^{\rm SM} = m_H^2/(2v^2)\approx 0.13$.
A promising candidate for an observable which is sensitive to $\lambda$
is double-Higgs boson production.
It has the potential to determine $\lambda$ with an uncertainty in the
10-20\% range in case LHC provides a luminosity of about $1000~{\rm fb}^{-1}$,
given that reliable theoretical predictions are at hand.

Leading order corrections to $gg\to HH$ have been computed quite some time ago
in Refs.~\cite{Glover:1987nx,Plehn:1996wb} where the exact dependence on all
kinematic variables has been taken into account. Next-to-leading order (NLO)
corrections in the effective theory approach where the top quark is integrated
out are known since about 15 years~\cite{Dawson:1998py}. A few months ago also
the NNLO corrections became available in this
limit~\cite{deFlorian:2013jea}\footnote{Note that the effective coupling of
  gluons to two Higgs bosons is still unknown.}  which works very well for
single-Higgs boson production.  The resummation of soft gluon radiation and
dominant $\pi^2$ terms have been considered in~\cite{Shao:2013bz} at
next-to-next-to-leading logarithmic order.  In Ref.~\cite{Grigo:2013rya} the
effect of a finite top quark mass has been considered at NLO. In this
contribution the findings of Ref.~\cite{Grigo:2013rya} are summarized.

\section{Technical details and LO result}

\begin{figure}[b]
  \begin{center}
    \includegraphics[width=0.95\columnwidth]{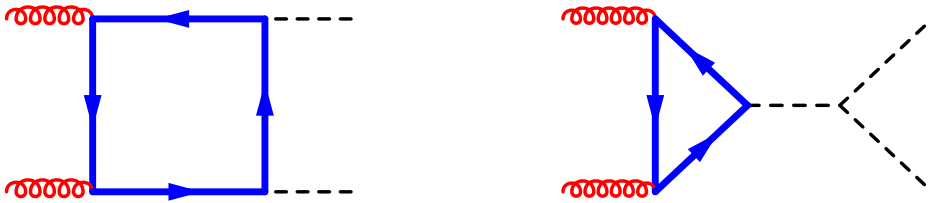}
    \caption{Box and triangle diagrams that contribute to double-Higgs boson
      production at leading order.  Solid lines refer to top quarks and
      dashed lines refer to Higgs bosons.  }
    \label{fig0}
  \end{center}
\end{figure}

\begin{figure}[t]
\begin{center}
  \mbox{}\quad\quad\quad\includegraphics[width=0.95\columnwidth]{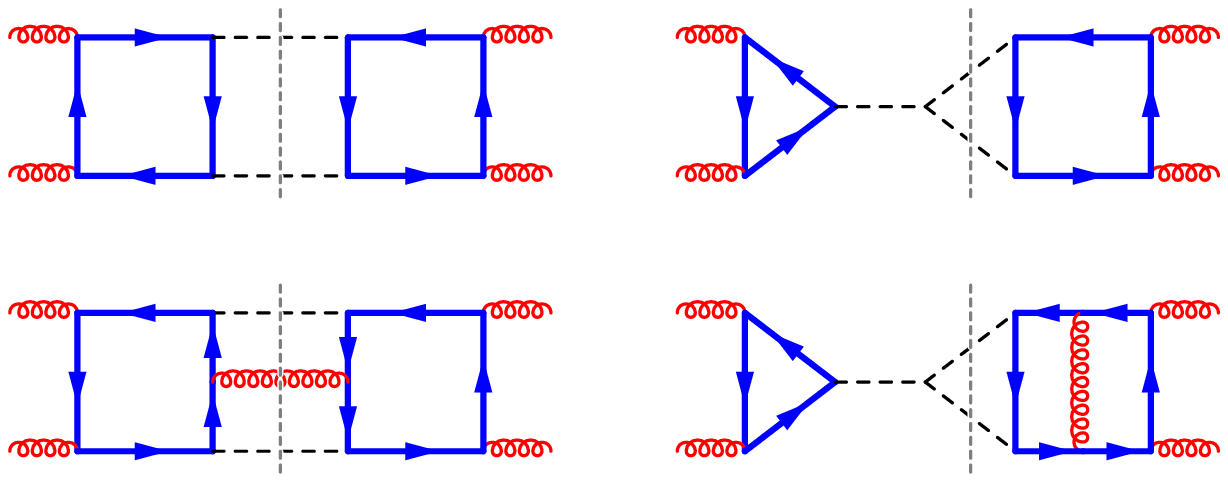}
  \\[3em]
  \includegraphics[width=0.95\columnwidth]{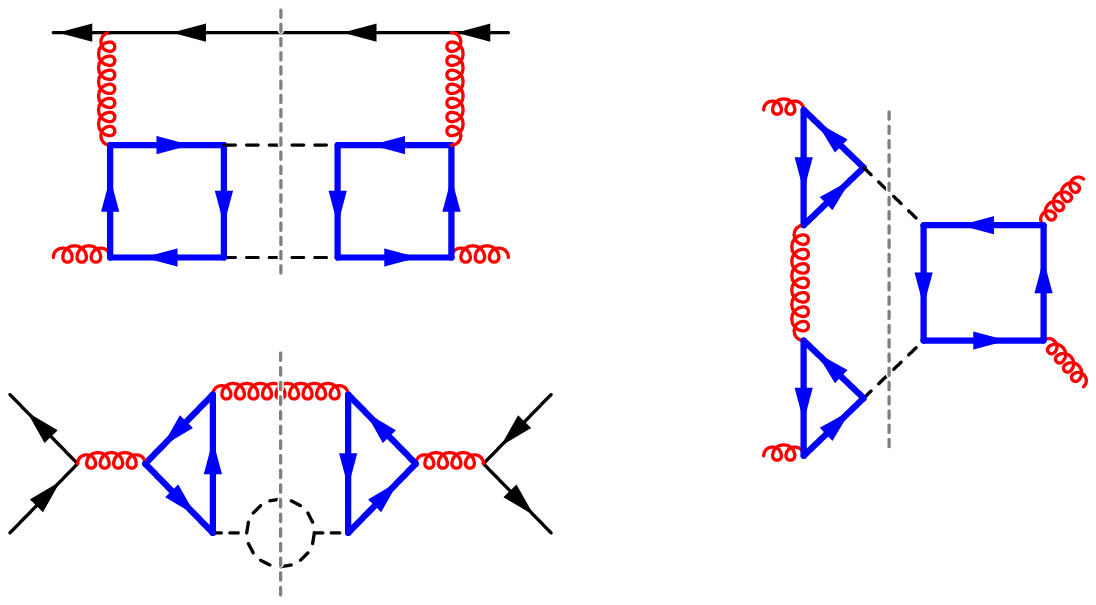}
  \\[2em]
  \caption{Sample Feynman diagrams for forward scattering kinematics which 
    contribute at LO (top row) and NLO.
    Dashed vertical lines represent unitarity cuts. Solid lines
    are top  and light quarks,  dashed  lines are Higgs bosons.}
\label{fig0a}
\end{center}
\end{figure}

The dominant production of Higgs boson pairs is loop-induced and
proceeds via gluon fusion. The contributing diagrams can be divided into a
triangle contribution, which has a dependence on the triple-Higgs boson
coupling and a box contribution where both Higgs bosons directly couple to the
top quark loop, see Fig.~\ref{fig0}. At NLO both virtual corrections and real
emissions have to be considered. In our approach they are incorporated via the
optical theorem which connects the total cross section to the imaginary part
of the forward scattering amplitude.  This has the advantage that, as pointed
out in Ref.~\cite{Anastasiou:2002yz}, well-established multi-loop techniques
can be applied to compute the Feynman integrals. Sample Feynman diagrams are
shown in Fig.~\ref{fig0a} where vertical dashed lines indicate the cuts which
lead to the imaginary part. On the right bottom corner of Fig.~\ref{fig0a} a
Feynman diagrams is shown which contains three closed top quark loops. Such
contributions appear for the first time at NLO and contribute to the virtual
part.  Fig.~\ref{fig0a} also contains Feynman diagrams illustrating the
quark-gluon and quark-anti-quark initiated channels which appear for the first
time at NLO. Note that they are suppressed by at least one order of magnitude
and thus in the remainder of this contribution we will restrict the discussion
to the gluon-gluon channel.

\begin{figure}[t]
  \begin{center}
    \includegraphics[width=0.8\columnwidth]{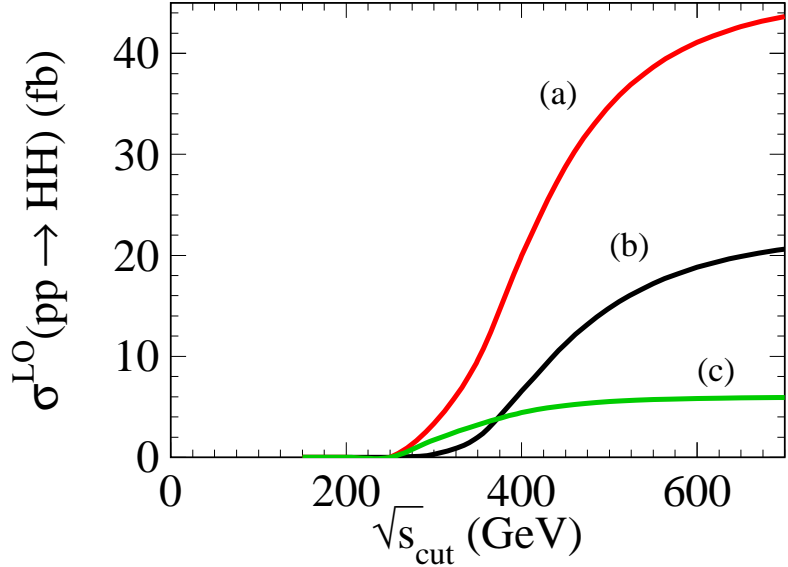}
    \caption{Leading order hadronic cross section for Higgs boson pair
      production at the $14~{\rm TeV}$ LHC as the function of the upper cut on
      the Higgs boson pair invariant mass.  Curve (b) is the full result;
      curve (a) is the box contribution; curve (c) is the triangle
      contribution.  The destructive interference between box and triangle
      contributions is apparent.  We use MSTW2008 parton distribution
      functions~\cite{Martin:2009iq}.  }
    \label{figm1}
  \end{center}
\end{figure}

The LO hadronic result for the Higgs boson pair cross section is shown in
Fig.~\ref{figm1} as a function of the invariant Higgs boson mass where curve
(b) corresponds to the SM. In order to demonstrate the sensitivity on the
triple-Higgs coupling it is switched off in curve (a) which leads to a
significantly higher cross section. For comparison we show in curve (c) the
result where the box contribution is set to zero. Although this part is much
smaller one observes a significant contribution from the interference
term. This is particularly true close to threshold where the SM result is
significantly smaller than the result from pure box or pure triangle
contribution.

The starting point of the NLO calculation are the four-loop diagrams in
Fig.~\ref{fig0a} which we evaluate in the limit where the top quark mass
is the largest scale. In that case the so-called hard-mass procedure (see,
e.g., Ref.~\cite{Smirnov:2013}) can be
applied which leads to a factorization of the integrals. To be precise, 
one obtains products of vacuum integrals up to two loops and
one- or two-loop integrals which depend on the Higgs boson mass
and the center-of-mass energy $\sqrt{s}$. The former are very well studied in
the literature whereas the latter have been considered for the first time in
Ref.~\cite{Grigo:2013rya}. We have organized our calculation in such a way
that in a first step the integrations associated with the massive vacuum
integrals are performed. Afterwards, we have computed integral tables
for the remaining one- and two-loop four-point functions with the help of
FIRE~\cite{Smirnov:2008iw,Smirnov:2013dia} which led us to four two-loop
master integrals. One of them can be expressed as a linear combination of the
others as described in Ref.~\cite{Grigo:2013rya}. For the remaining 
three diagrams an integral representation can be derived which can easily be
Taylor-expanded in the parameter $\delta = 1 - 4m_H^2/s$. Note that $\delta$
vanishes at threshold. We have computed 100 expansion terms for the partonic cross
section and
have checked that there is no change in the hadronic cross section in case 
less terms are taken into account, e.g., only terms up to order $\delta^{50}$.
In principle the master integrals could also be evaluated numerically,
however, in our approach analytic results for the partonic cross section can
be provided.

\section{NLO result}

\begin{figure}[t]
  \begin{center}
    \includegraphics[width=0.8\columnwidth]{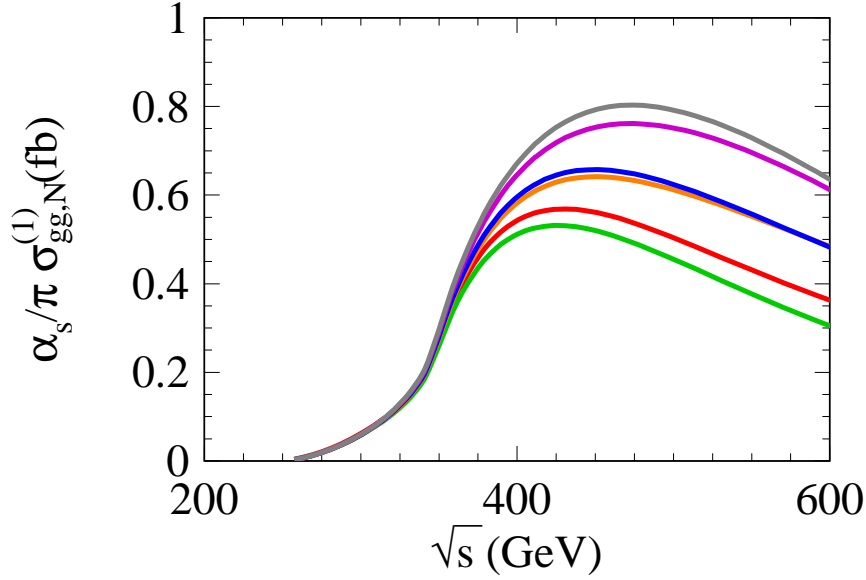}
    \caption{Next-to-leading order contribution to $gg \to HH$
      re-scaled by the exact leading order result
      where the curves from bottom to top corresponds to
      $N=1,0,2,3,4,5$.
      For the Higgs boson and top quark mass we have used
      $m_H = 126~{\rm GeV}$ and $M_t = 173.18~{\rm GeV}$.}
    \label{fig::ggpart}
  \end{center}
\end{figure}

Using the approach described in the previous section we have computed
six\footnote{At the time Ref.~\cite{Grigo:2013rya} has been published only
  five terms were available.}
expansion terms of the partonic cross section in $\rho=m_H^2/M_t^2$.
On general grounds we expect that such a series shows good convergence
properties below the threshold where two real top quarks can be produced,
i.e., for $\sqrt{s}\lsim 346$~GeV. This is observed in practice. At the same
time higher order expansion terms provide sizeable corrections above
threshold. In analogy to single-Higgs boson production it is possible
to improve the convergence by factoring out the exact leading
order cross section and by defining the NLO contribution via
\begin{eqnarray}
  \sigma_{ij,N}^{(1)} = 
  \sigma_{gg,\rm exact}^{(0)}  \Delta_{ij}^{(N)},
  \qquad \Delta_{ij}^{(N)} = 
  \frac{\sigma_{ij,\rm exp}^{(1)}}{\sigma_{gg,\rm exp}^{(0)}} = 
  \frac{\sum \limits_{n=0}^{N} c_{ij,n}^{\rm NLO} \, \rho^n}{\sum
    \limits_{n=0}^{N} c_{gg,n}^{\rm LO} \, \rho^n}
  \,,
  \label{eq::nlo}
\end{eqnarray}
where terms up to $\rho^N$ are included in the approximation.
Note that the quantity $\Delta_{ij}^{(N)}$ in Eq.~(\ref{eq::nlo})
is a rational function in $\rho$ which is crucial in order to obtain good
results.

In Eq.~(\ref{eq::nlo}) the complete LO cross section consisting of triangle
and box contributions has been factored out. Alternatively, one could also
separate the NLO expression into triangle-triangle, box-box and triangle-box
contributions and factor out for each piece the corresponding exact Born cross
section. We have implemented also this option and checked that on the
practical level there is no difference as compared to the approximation of
Eq.~(\ref{eq::nlo}).

In Fig.~\ref{fig::ggpart} we show the cross section as a function of the
partonic center-of-mass energy $\sqrt{s}$.  Reasonable convergence is observed
up to $\sqrt{s}\approx 400$~GeV but also for higher energies the difference
between the curves with $N=0$ and $N=5$ remains bounded. Since the weight of
this region is suppressed in the convolution integral with the parton distribution
functions we expect even better results for the hadronic cross section.

\begin{figure}[t]
  \begin{center}
    \includegraphics[width=0.8\columnwidth]{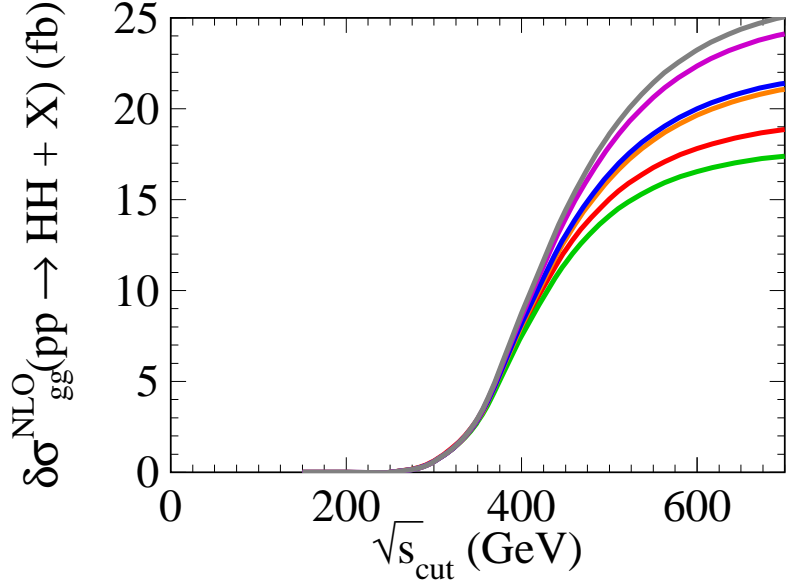}
    \caption{Next-to-leading order contributions for the gluon-gluon channel
      to the hadronic production cross section of Higgs boson pairs. 
      From bottom to top the curves correspond to $N=1,0,2,3,4,5$.}
    \label{fig::gghadr}
  \end{center}
\end{figure}

Fig.~\ref{fig::gghadr} shows the hadronic gluon-induced cross section for
$N=0,\ldots,5$ as a function of $\sqrt{s_{\rm cut}}$, an upper cut on the
partonic center-of-mass energy. At LO $\sqrt{s_{\rm cut}}$ is equivalent 
to the two-Higgs boson invariant mass $M_{HH}$. At NLO this is not true
any more, however, we use $\sqrt{s_{\rm cut}}$ as an approximation to the cut
on $M_{HH}$. Note that in the limit $\sqrt{s_{\rm cut}}\to \infty$ the total 
cross section is recovered. In particular, for $N=0$ the results of
Ref.~\cite{Dawson:1998py} could be confirmed.\footnote{Note that in
  Ref.~\cite{Dawson:1998py} the factorization of the exact LO result has been
  implemented differently from Ref.~\cite{Grigo:2013rya}.}

\begin{table}[b]
  \begin{center}
    \begin{tabular}{l||l|l|l|l|l|l||l}
      $\sigma^{\rm LO}$ (fb) & 
      \multicolumn{6}{c||}{$\delta\sigma^{\rm NLO}_N$ (fb)} & 
      $\sigma^{\rm NLO}$ (fb) \\
      \hline
      & N=0 & 1 & 2 & 3 & 4 & 5 & \\
      \hline
      22.4 & 
      19.0 & 16.4 & 21.5 & 21.4 & 24.5 & 25.3 &
      45.0 $\pm$ 3.9
    \end{tabular}
    \caption{LO and NLO cross sections for double-Higgs boson production
      including also the quark channels at NLO. No cut on the partonic
      center-of-mass energy is applied. For $\sigma^{\rm LO}$ the LO
      MSTW2008 parton distribution functions~\cite{Martin:2009iq} have been
      used and all other results have been obtained using NLO parton
      distribution functions. We have set the hadronic center-of-mass energy
      to 14~TeV.}
    \label{tab::xs}
  \end{center}
\end{table}

It is interesting to mention that the expansion terms in
Fig.~\ref{fig::gghadr} group into pairs which contain two successive orders. 
Note that for $\sqrt{s_{\rm cut}} \sim 700~{\rm GeV}$, the shift
in $\delta \sigma_{gg}^{\rm NLO}$ due to the last computed term in the $1/M_t$
expansion as compared to the prediction including $\rho^3$ terms
is less than 20\%; for lower values of $\sqrt{s_{\rm cut}}$
the convergence is significantly better. Taking into account that the $K$
factor is close to two these shifts get basically reduced by the same factor
when considering the complete NLO prediction.

We refrain to discuss the $K$ factor in detail in this contribution and refer
to Ref.~\cite{Grigo:2013rya}. Instead we want to provide results for the
total cross section obtained after removing the cut on the partonic
center-of-mass energy. In Tab.~\ref{tab::xs} we summarize the results
of $\sigma^{\rm LO}$, $\delta\sigma^{\rm NLO}_N$ and 
$\sigma^{\rm NLO} = \sigma^{\rm LO}+\delta\sigma^{\rm NLO}_5$ 
where the subscript $N$ indicates the used expansion depth in $\rho$ for the
partonic cross section. The uncertainty assigned to $\sigma^{\rm NLO}$
has been obtained by comparing $\delta\sigma^{\rm NLO}_5$ and
$\delta\sigma^{\rm NLO}_3$ which is a quite conservative approach.

\section{Concluding remarks}

\begin{figure}[t]
  \begin{center}
    \includegraphics[width=0.8\columnwidth]{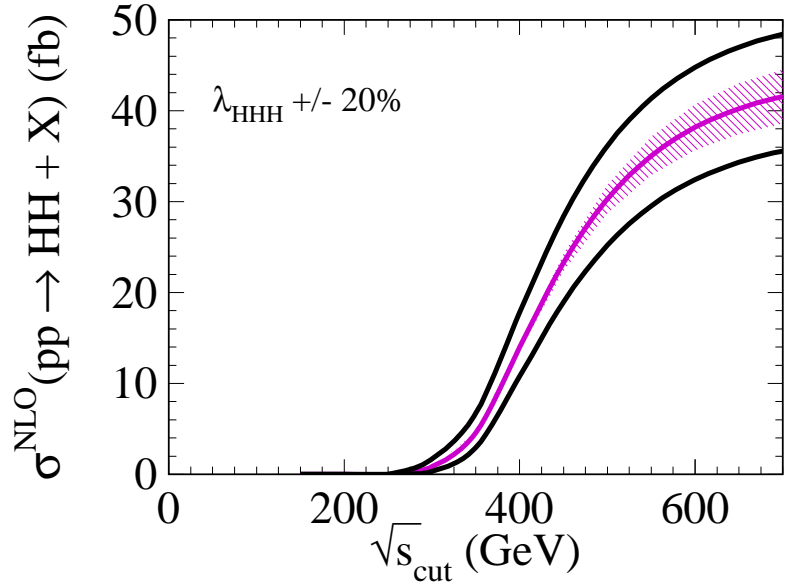}
    \caption{The NLO hadronic cross section at the $14~{\rm TeV}$ LHC as a
      function of $\sqrt{s_{\rm cut}}$.  Two black curves correspond to $\pm
      20\%$ variation in the triple-Higgs boson coupling relative to its SM
      value.}
    \label{fig::hadr_all}
  \end{center}
\end{figure}

This contribution summarizes the computation of top quark mass effects to the
cross section $\sigma(pp\to HH)$ at NLO~\cite{Grigo:2013rya}.  An expansion
for a heavy top quark mass has been applied and six expansion terms have been
computed. After factoring out the exact LO result at partonic level one
observes a good convergence of the hadronic cross section close to threshold,
the energy region most important for the extraction of the triple-Higgs boson
coupling.  Thus, for the first time it is possible to quantify the importance
of finite top quark mass effects. This is illustrated in
Fig.~\ref{fig::hadr_all} where the NLO cross section is shown as a function of
$\sqrt{s_{\rm cut}}$.  The (violet) uncertainty band due to uncalculated $1/M_t$
corrections has been obtained by comparing the NLO results including $\rho^4$
and $\rho^3$ terms which amounts to about 10\%.  Furthermore, we vary the
triple-Higgs boson coupling relative to its SM value by $\pm 20\%$ which is
shown as black curves in Fig.~\ref{fig::hadr_all}. This leads to the
conclusion that the current knowledge of $1/M_t$ corrections is sufficient to
be sensitive to detect ${\cal O}(10 \%)$ deviations in the triple-Higgs boson
coupling, relative to its SM value.

Results for the total cross section can be found in Tab.~\ref{tab::xs}. The
NLO contributions increase by about 30\% when going from the infinite-top
quark mass limit to the result including $\rho^5$ terms. 
Furthermore, we can quantify the uncertainty at NLO due to the lack of 
exact $M_t$ dependence which amounts to
about 9\% for the NLO prediction of the total cross section.

\section*{Acknowledgments}

The research of J.H, J.G. and M.S. is supported by the Deutsche
Forschungsgemeinschaft in the Sonderforschungsbereich Transregio~9
``Computational Particle Physic''.  The research of K.M. is partially
supported by US NSF under grants PHY-1214000.  The research of
K.M. and J.G. are partially supported by Karlsruhe Institute of
Technology through its distinguished researcher fellowship program.

\end{document}